# Assessment and support of emerging research groups

Henk F. Moed

Sapienza University of Rome, Italy. Email: hf.moed@gmail.com

## Abstract

The starting point of this paper is a desktop research assessment model that does not take properly into account the complexities of research assessment, but rather bases itself on a series of highly simplifying, questionable assumptions related to the availability, validity and evaluative significance of research performance indicators, and to funding policy criteria. The paper presents a critique of this model, and proposes alternative assessment approaches, based on an explicit evaluative framework, focusing on preconditions to performance or communication effectiveness rather than on performance itself, combining metrics and expert knowledge, and using metrics primarily to set minimum standards. Giving special attention to early career scientists in emerging research groups, the paper discusses the limits of classical bibliometric indicators and altmetrics. It proposes alternative funding formula of research institutions aimed to support emerging research groups.

## Introduction

During the past decades, major changes took place in the domain of research assessment, under the influence of developments in science policies, information and communication technologies, and in quantitative assessment methodologies. In many OECD countries national governments seek to increase both the effectiveness and the efficiency of government-supported research. Therefore, they need systematic evaluations of their research systems tooptimize their research funding, often in large scale research assessment exercises (OECD, 2010; Hicks, 2010).

Especially in universities, government funding of scientific research is increasingly based upon performance criteria. As research institutions operate more and more in a global market, international comparisons of institutions are published on a regular basis. The primary aim of such ranking systems is to inform students, researchers and knowledge seeking external groups. Typical examples are the Shanghai, Times Higher Education, QS and Leiden Ranking, U-Multirank, and Scimago Country and Institute Rankings. Research managers use this information to benchmark the performance of their institutions against that of their competitors (Hazelkorn, 2011). At the same time, institutions build research information systems and implement internal research assessment processes (see for instance EUROCRIS, 2013)

The digitization of scholarly communication enables the creation of large, policy-relevant data sources on research output and impact, making quantitative research assessment look more and more like a 'big data' science. Nowadays, multiple, comprehensive citation indexes have been created to monitor research trends at a global level. More and more publication archives do not only contain publication meta-data, but also full texts of scientific-scholarly publications in digital format.

While the digitization of scholarly communication created information systems on 'usage' – i.e., on downloading and online browsing – of scientific publications, researchers' use of social media such as Twitter, reference managers like Mendeley, and scholarly blogs, to communicate with each other and to promote their work leaves behind traces that can be analysed with informetric techniques. This is also true for the way in which institutions manifest themselves in their websites on the internet.

The above trends have generated an increasing interest in the development, availability and practical application of new indicators for research assessment. Newly developed indicators cover new dimensions of research communication and performance. Typical examples are altmetrics, webometrics and usage based measures, but also citation-context analyses, topic mapping and author network analysis. Their development attracted specialists from many other disciplines, especially from statistical physics, molecular biology, econometrics, computer science and digital humanities.

Currently, indicators such as author publication counts, total citation counts and h-indices, as well as journal citation metrics, are available in the three large literature databases, Web of Science, Scopus and Google Scholar, or in special assessment tools such as Clarivate's Incites, Elsevier's SciVal, or Digital Science's Dimensions. Other metrics are produced by small, specialized firms. Typical examples are Altmetric.com, Plum Analytics, and service companies closely linked to academic groups, such as Scimago.com and CWTS.com. Platforms such as Mendeley and ResearchGate provide indicators as well, based on data from their own systems.

It is clear that both the calculation and the interpretation of science metrics are not merely made by bibliometric experts. "Desktop bibliometrics", a term coined by Katz and Hicks (1997) and referring to an evaluation practice using bibliometric data– in a manner of speaking – by pushing a button, and based on simplifying, questionable assumptions on what the indicators measure, is pervading the assessment domain.

At the same time, researchers express more and more *critique* on the actual use of indicators in assessment processes. For instance, Benedictus and Miedema (2016) argued that a pressure to publish as many papers and generate as many citations as possible, created in their view in the Dutch academic system a tendency to give too little importance to the effect of research work on patient care. Their experiences seem to be illustrations of the constitutive effects of indicators expressed by Dahler-Larsen (2013). The notion of desktop bibliometrics is further developed below.

*The line of argument in the current paper*

The current paper focuses on formalized assessment processes, conducted within the framework of decisions on hiring or promotion of individual researchers, and on performance based funding of research departments and institutions. Its starting point is a desktop research assessment model. This model embodies a series of simplifying assumptions related to the availability, validity and evaluative significance of research performance indicators, the ordering principle of units under assessment, and the criteria on which a policy decision about 'support' is based. The next section gives a more detailed description of this model. The paper defends the position that, although this model may, at first sight, have a certain plausibility, a reflection reveals that the validity of its assumptions is questionable, and that alternative assumptions are defensible and worthwhile being seriously considered.

The next sections present a critique on the desktop model, and propose alternative approaches. Firstly, an analytical distinction is made between four domains of intellectual activity in research assessment, namely policy, evaluation, analytics and data collection. It provides a theoretical background for the discussion of the pros and cons of the various types of metrics. A key notion is that evaluative informetrics itself does not evaluate, and is based on values that cannot be grounded in informetric research. It is one of the central theses defended by Moed (2017). Therefore, in any assessment an evaluative framework is needed that specifies the evaluation criteria. In many applications such a framework is implicit.

A next section discusses the merits and limits of a number of important bibliometric or informetric indicators It starts with a discussion of three widely available bibliometric indicators, namely publication counts, journal impact factors and h-indices. The first two measures were introduced more than 50 years ago, almost simultaneously with the foundation of the Science Citation Index. Although the third is of a more recent date (2005), these three indicators will be denoted as 'classical'. Next, the potential is discussed of new measures denoted as altmetrics, focusing on indicators derived from social media and full text downloads.

A series of alternative assessment approaches and funding formula in the last section, following suggestions made by Moed (2017, Ch. 10). First, it highlights a series of general principles, focusing on preconditions to performance or communication effectiveness rather than on performance itself, and on using metrics primarily to set minimum standards rather than focus on the top of indicator-based rankings. Next, it gives special attention to a most important category of researchers, namely early career researchers (ECRs). The value of the suggested approaches in the assessment of early career scientists and emerging groups is illustrated. An alternative funding formula of research institutions is suggested that supports ECRs more strongly than other performance-based funding formula do.

*Early career researchers and emerging research groups*

Nicholas et al. (2018) define early career scientists (ECRs) as follows. "ECRs are generally not older than thirty-five and have either received their doctorate and are currently in a research position or have been in research positions and are currently pursuing a doctorate. In neither case are they researchers in established or tenured positions. In the case of academics, they are non-faculty research employees of the university".

The current paper gives special attention to the class of early career researchers who have received their doctorate and have been active for some years in a postdoctoral research position, and who have the ambition to further develop as researchers, by obtaining a more permanent or tenured position. It acknowledges that in most if not all disciplines research is not conducted alone, but in teams. This team can be small and consist of two members only, who have a master-apprentice relationship, or be larger, and include colleague PhD or Postdoctoral students, other seniors and technical staff.

As ECRs are normally trained within a team, their past performance depends upon team work, and so does their future when they aim to build up a research group centred around a program in which they have acquired advanced know-how. Such a group is denoted below in as an emerging research group. This emphasis on group formation parallels the notion that the assessment of an individual involves the task to value his contribution to the creation of a common research product, being a multi-authored publication or another type of scholarly output, an issue that is further discussed below.

## The desktop assessment model.

The characteristics of this model are summarized in the second column in Table 1. The last column presents the basic notions of the alternative indicators, assessment approaches and funding formula further discussed in later sections.

Table 1. The desktop assessment model and alternative approaches

|   | Aspect | Assumptions of the Desktop Model | Alternative approaches |
|---|---|---|---|
| 1 | Availability | Widely available indicators should be used (i.c., publication counts, journal impact factors, h-indices) | Use tailor-made indicators appropriate for the application context |
| 2 | Validity | Indicators measure what they are supposed to measure well enough; no confirmation from other sources is needed | Indicators should be combined with expert knowledge and measure pre-conditions of performance rather than performance itself |
| 3 | Evaluative significance | The aspects measured by the indicators constitute appropriate evaluation criteria | An independent evaluative framework is needed |
| 4 | Ordering principle | The higher the score, the better the performance | Consider using indicators to define minimum standards rather than identifying the top of a ranking |
| 5 | Policy decision criteria | The best performer receives the largest support | Support ECSRs by funding institutions according to the number of emerging research groups |

The following explanatory comments should be made. Firstly, the wide availability of publication counts, journal impact factors and h-indices is mainly due to the fact that three large citation indexes, namely Web of Science (Clarivate Analytics), Scopus (Elsevier) and Google Scholar (Google), offer for large numbers of individual authors publication counts and h-indices – although the data underlying these indicators are often not verified by the authors themselves, and may therefore be inaccurate – and several types of journal citation impact indicators for a large set of periodicals.

It must be underlined that the fact that indicators are widely *available* does *not* imply that they are often *used*. In this respect the study by Nicolas et al. (2018) is informative who found that most of the about 100 ECRs from various countries interviewed in the study indicated that either they have been subjected to this model themselves, or that they *believed* it was the dominant model in their institution of country. More empirical research is needed in the way in which performance indicators are actually being used in research assessment processes.

A base assumption of the desktop model is that if one unit of assessment has a higher score on a particular indicator than a second unit, this provides an indication that the first has performed better the second. As a consequence, if the unit has the largest score in a wider set, this indicates that it is the best performer of all units in that set. This assumption almost automatically seems to focus the attention to the top of the quality distribution.

The validity of an indicator relates to the issue as to whether it adequately measures the concept it claims to measure, e.g., do citations measure research quality? In the desktop model this issue is discarded as an unproblematic issue that has been solved by scientometric experts. On the other hand, scientometric experts themselves have underlined in numerous papers and presentations that in order to draw valid conclusions about the performance of a unit if assessment, indicators should be combined with expert knowledge on this unit (e.g., Van Raan, 2004: Moed, 2005; Hicks et al., 2015).

The evaluative significance refers to the question whether a particular concept constitutes an adequate assessment criterion. For instance, is quality or impact what needs to be evaluated? The next section further discusses the role of an evaluative framework. If the evaluation criteria are vague and defined in very general terms (e.g., "research quality"), assessors may be more inclined to use indicators developed in a particular context in other contexts as well, and may not realise sufficiently well that the choice of indicators strongly depends upon the object of assessment, the aspects to be evaluated, and its policy context and objectives (e.g., Moed, 2017, Ch 8).

As a final comment, in de desktop model, it is assumed that the amount of funding should depend positively on the indicator score. In short, the higher the indicator score, the larger should be the amount of support. Although this principle may at first sight have a certain plausibility, Bishop (2013) and other authors have shown that its application at a national scale for the distribution of funds among universities causes at the longer large disparities among institutions, while its effects upon the performance of the system as a whole are at least uncertain.

## Four domains of intellectual activity in research assessment

A distinction should be made into four domains of intellectual activity in a research assessment process. Table 2 summarizes their main characteristics. Moreover, it gives typical examples of aspects and issues in the various domains that are derived from the current standard application model.

Table 2. Four domains of intellectual activity in an assessment process according to Moed, 2017, Ch 6.

| Level | Key aspect and issues | Main outcome |
| --- | --- | --- |
| Policy | Desirable objectives; strategies to serve them; Implementation of the assessment. Are objectives and strategies fair and aligned with the rules of good governance? | Policy decision based on the outcomes from the evaluation domain |
| Evaluation | Evaluative framework: what is valuable and how is value to be assessed. What is it that constitutes performance, and do the indicators capture it adequately? | A *judgment* on the basis of an evaluative framework and the empirical evidence collected. |
| Analytics | Empirical and statistical research: development of new methods; indicator validity; effectiveness of political strategies. | An analytical report as input for the evaluative domain. |
| Data collection | Creation of databases with data relevant to the analytical framework; data cleaning; assessment of data quality. | A dataset for the calculation of all indicators specified in the analytical model. |

Informetric analysis is to be positioned in the analytics domain. The monograph Applied Evaluative Informetrics by Moed (2017) defends the following position.

"Evaluation criteria and policy objectives are not informetrically demonstrable values. Of course, empirical informetric research may study quality perceptions, user satisfaction, the acceptability of policy objectives, or effects of particular policies, but they cannot provide a foundation of the validity of the quality criteria or the appropriateness of policy objectives. Informetricians should maintain in their informetric work a neutral position towards these values. (Moed, 2017, p. 12)".

The final section presents typical examples of evaluative frameworks related to emerging research groups.

## Pros and cons of classical indicators and altmetrics

As a starting point, the current section presents in a critical discussion of an assessment model using one or more of three widely available bibliometric indicators, namely publication counts, journal impact factors and h-index. The first two measures were introduced more than 50 years ago, almost simultaneously with the foundation of the Science Citation Index. Although the third is of a more recent date (2005), these three indicators will be denoted as 'classical'. Next, the potential is discussed of a series of new measures denoted as altmetrics and based on data derived from social media.

*Publication counts.*

Publications constitute in all scientific-scholarly subject fields an important type of output. Although mostly applied to journal articles, publication counts may also include books and other types of textual output. They constitute a useful tool to identify, at the bottom of the performance distribution, entities (groups, departments, institutions) that are not sufficiently active in research, to the extent that their number of published papers lies below a certain – subject field dependent – *minimum*. If the publication counts exceed this minimum level, differences between them cannot be interpreted in terms of performance. Moed summarizes the limits of publication-based indicators as follows.

- "If numbers exceed a certain minimum level, differences between them cannot be interpreted in terms of performance.
- Despite large differences in the level of publication output between subject fields, publication counts are not field-normalized.
- They have a limited value in the private sector, and in technical science.
- Collaboration and multi-authorship is a rule rather than an exception. How to assess the contribution of an entity to the output of team work?" (Moed, 2017, p. 51).

The latter issue is especially relevant in the case of the assessment an individual researcher who has published a large fraction of his or her publications jointly with other authors.

*Journal impact factors.*

The journal impact factor (JIF) introduced by Eugene Garfield was created as a tool (combined with expert opinion) in the selection of source journals for the Science Citation Index, *independently* of scientific publishers. Later, the JIF was more and more used to assess the performance of individuals and groups, based on the JIF values of the journals in which they published. On the pro side, the quality or impact of the journals in which an individual or group under assessment has published is a performance aspect in its own right. Moreover, JIF is a simple measure; is already used for decades by researchers, librarians and publishers; and is available for large numbers of journals. But JIF has several technical limitations as well. Firstly, it does not correct for differences in levels of citation frequency between subject fields. Next, a well known technical problem is that for many journals discrepancies arise between document types included in the impact factor's numerator and those counted in its denominator, leading to so called "free" citations. It has been shown that this measure can be strongly affected by so called citation outliers, and biased in favour of journals publishing a large fraction of review articles.

Newly proposed journal citation measures do not suffer from at least some these problems. But even the application of these alternative measures is far from problematic. Journal metrics are no good predictors of the actual citation rate of individual papers; moreover, their values can to some extent be manipulated and may be affected by particular editorial policies. One of the justifications of their use is that a journal's citation impact correlates positively with the rigorousness or 'quality' of its manuscript peer review process. But the current author defends the position that there is no sound empirical basis for this claim (Moed, 2017, Ch 12.2).

*H-index*

The creator of the h-index claimed that "this index may provide a useful yardstick to compare different individuals competing for the same resource when an important evaluation criterion is scientific achievement, in an unbiased way" (Hirsch, 2005). On the one hand, this indicator combines an assessment of both quantity (published papers) and impact (citations), and tends to be insensitive to highly cited outliers and to poorly cited papers. But on the other hand, its value is biased in favour of senior researchers compared to juniors – for instance, the h index cannot exceed the number of publications made – and does not correct for differences between subject fields.

Hirsch was well aware of both limitations. As regards the first, he introduced a second parameter, *m*, that could be used to correct for differences in academic age. Although Google Scholar, Scopus and Web of Science publish h-indices for large numbers of authors, they do not present any data on this second parameter. Hirsch proposed specific threshold values for advancement to tenure, full professor and scholarly society member; for instance, h-index values of 10-12 could be typical values of advancement to tenure. Despite the creativity of the indicator concept, the current author defends the position that such an algorithmic approach to career advancement of individuals suggests a false precision.

*Full text downloads*

On the pro side, full text download counts of an article are in principle available immediately after its online publication. They enable researchers to assess the effectiveness of their communication strategies, and may reveal attention of scholarly audiences from other research domains or of non-scholarly audiences. But a series of technical and validity issues makes them as of yet of little use of fully-fledged metrics in performance assessments.

Moed (2017) summarizes the limitations of the use of usage-based indicators as follows:

- "Downloaded articles may be selected according to their face value rather than their value perceived after reflection.
- Incomplete availability of usage data across providers.
- Usage counts are affected by differences in reading behaviour between disciplines and institutions.
- Usage counts can be manipulated.

- It is difficult to ascertain whether downloaded publications were actually read or used" (Moed, 2017, p.55).

Kurtz and Bollen (2010) denote download counts as measures of *attention*. They underline that receiving attention is in itself not a sufficient basis for a performance measurement. After all, the principle objective of research is to contribute to scientific-scholarly progress or to the solution of societal problems, not to create attention as such.

*Social media mentions*

Strong points of the use of mentions of publications in Twitter, Facebook, blogs, and other social media is that these data are immediately available after publication, and may potentially reveal impact upon scholarly audiences from other research domains or upon non-scholarly audiences; they may also provide tools to link scientific expertise to societal needs. But Moed defends the position that "scientific-scholarly impact and societal impact are distinct concepts", and that "one cannot measure scientific-scholarly impact with metrics based on social media mentions"(Moed, 2017, p. 55).

Thelwall (2014) concludes that altmetrics may have a certain value as indicators of research impact, to the extent that they provide insights into aspects of research that were previously difficult to study, such as the extent to which articles from a field attract *readerships* from other fields or the value of social media publicity for articles. Altmetrics also have the potential to be used as impact indicators for individual researchers. But he warns that "this information should not be used as a primary source of impact information since the extent to which academics possess or exploit social web profiles is variable (Thelwall, 2014)". He also underlines that altmetrics are easily manipulated, "since social websites tend to have no quality control and no formal process to link users to offline identities" (Thelwall, 2014). Haustein (2016) presents a good overview of the heterogeneity, data quality and dependencies of altmetric data.

Indicators derived from reference managers and reader libraries such as Mendeley and ResearchGate deserve special attention (see for instance Thelwall & Kousha, 2014). Manuscripts in preparation are potentially predictors of emerging trends and future citation impact. But the outcomes of their analysis depend on readers' cognitive and professional background and are not necessarily representative of the global research community. Also, it seems unthinkable that a measure as little transparent as RG Score in ResearchGate could play a serious role in a performance assessment of individual researchers or groups.

On the one hand, there are high expectations of these indicators: proponents claim that they may reflect performance in a better or at least a more comprehensive manner than the classical indicators mentioned above. But current experiences suggest that they do not – or perhaps not yet – provide useful tools to performance assessments. This is not so much because practitioners are not accustomed to measuring performance in this manner, but rather because there are serious methodological problems and shortcomings that seem to make them unusable for this end.

## Alternative approaches related to early career scientists

This section starts with making a series of general proposals for alternative approaches toward metrics-based assessment of research performance, presented in Moed (2017, Ch 10). They have direct consequences for the way in which ECRs are evaluated. The second part focuses on ECRs. Rather than focusing on individual researchers, it deals with emerging research groups, and discusses indicators to identify them, and funding mechanisms to support them.

*General characteristics*

Moed (2017) proposes the following alternative approaches to the assessment of academic research.

- "A key assumption in the assessment of academic research has been that it is not the potential influence or importance of research, but the actual influence or impact that is of primary interest to policy makers and evaluators. But an academic assessment policy is conceivable that rejects this assumption. It embodies a shift in focus from the measurement of performance itself to the assessment of preconditions for performance.
- Rather than using citations as indicator of research importance or quality, they could provide a tool in the assessment of communication effectiveness, and express the extent to which researchers bring their work to the attention of a broad, potentially interested audience. This extent can in principle be measured with informetric tools. It discourages the use of citation data as a principal indicator of importance.
- The functions of publications and other forms of scientific-scholarly output, as well as their target audiences should be taken into account more explicitly than they have been in the past. Scientific-scholarly journals could be systematically categorized according to their function and target audience, and separate indicators could be calculated for each category. More sophisticated indicators of internationality of communication sources can be calculated than the journal impact factor and its variants.
- One possible approach to the use of informetric indicators in research assessment is a systematic exploration of indicators as tools to set minimum performance standards. Using baseline indicators, researchers will most probably change their research practices as they are stimulated to meet the standards, but if the standards are appropriate and fair, this behavior will actually increase their performance and that of their institutions" (Moed, 2017, p. xii).

*Early career scientists and emerging groups*

The term evaluative framework introduced above refers to a specification of the qualitative principles and considerations that provide guidance to a *concrete assessment process*. A core element in an evaluative framework is the specification of a *performance criterion*, in a set of propositions on what constitutes research quality or performance. Such values cannot be grounded in bibliometric/informetric research. But from such propositions follow the indicators that should be used, and, in a next logical step, the data sources from which these are to be calculated.

Below follow typical examples of possible evaluative frameworks. They illustrate that the choice of indicators – as well as the underlying performance criterion – depend strongly upon the context: what is the unit of assessment; which quality dimension is to be assessed; what is the objective of the process? And what are relevant, general 'systemic' characteristics of the units of assessment?

*Example 1: emerging research groups*

Several bibliometric groups, especially the group at CWTS (e.g., Moed et al., 1985) have developed a bibliometric method aimed to accommodate a particular policy need, namely the development of tools for identifying scientifically *emerging* and *declining* research groups. Indicators should, for instance, be able to assign to a small emerging group with a relatively short track record and a still limited level of external, competitive funding, but already rapidly increasing its international visibility, a higher score than a large department that had consolidated its status for many years, following research lines in which international colleagues in the field were losing interest. The use of so-called 'size normalized' indicators seemed appropriate, especially the trend in a short-term citation-per-article indicator, taking into account differences in citation practices between subject fields. Total publication or citation counts, and also journal impact factors were considered inappropriate indicators in the given context. Nor would the h-index, launched 20 years after the CWTS study, be a useful measure.

An emerging research group was described as a group centred around a research program in which they have acquired advanced know-how. This mean that one would expect there to be one or a few 'key publications' written by the group leader as primary author, publications that mark the intellectual base of the program, and are cited frequently relative to other articles of about the same age in the same subject. A citation analysis could provide an indication of the existence and the status of such papers. But given the biases inherent to citation analysis, especially at the level of individuals and small groups, expert and background knowledge on the group leaders' performance, including his or her group-leading capabilities, are indispensable. Therefore, the assessment would combine metrics and peer review.

*Example 2: alternative funding formula*

In response to major criticisms towards current national research assessment exercises and performance-based funding formula, an alternative model is conceivable that would require less effort, be more transparent, stimulate more strongly new research lines and reduce to some extent the 'Matthew Effect', according to which 'the rich get richer'. The basic unit of assessment in such a model is the emerging research group rather than the individual researcher. Institutions submit emerging groups and their research programs, which are assessed in a combined peer review-based and informetric approach, applying minimum performance criteria. A funding formula could be partly based on an institution's number of 'acknowledged' emerging groups. This funding would not be given for the emerging group only, but more as a kind of block grant to institutions as a premium to their supporting emerging research groups.

Evidently, the examples of alternative approaches to assessment and funding of research present only the main lines. Many issues still have to be addressed, and more details have to be specified. But at least the examples indicate a direction of thinking, and illustrate that alternative approaches are conceivable, especially for early career scientists. Much more work has to be done to develop assessment models that profit from the advantages of performance indicators, while reducing their biases and taking into account their limits. But in the view of the current author quantitative research assessment methodologies do have a great potential.

## Acknowledgement

The author is grateful to two anonymous reviewers for their valuable comments on an earlier version of this paper.

_